\documentclass[sn-mathphys-num,iicol,twocolumn]{sn-jnl}

\usepackage{graphicx}%
\usepackage{multirow}%
\usepackage{amsmath,amssymb,amsfonts}%
\usepackage{amsthm}%
\usepackage{mathrsfs}%
\usepackage[title]{appendix}%
\usepackage{xcolor}%
\usepackage{textcomp}%
\usepackage{manyfoot}%
\usepackage{booktabs}%
\usepackage{algorithm}%
\usepackage{algorithmicx}%
\usepackage{algpseudocode}%
\usepackage{listings}%
\usepackage{hyphenat}

\theoremstyle{thmstyleone}%
\theoremstyle{thmstyletwo}%

\theoremstyle{thmstylethree}%

\begin{document}

\title[Article Title]{HARPPP: Autonomous Geometric Design Optimisation of Stirred Tank Reactor Impellers and Baffles}


\author*[1]{\fnm{A. Leonard} \sur{Nicușan}}\email{leonard@evophase.co.uk}

\author[2]{\fnm{Darren} \sur{Gobby}}\email{darren.gobby@matthey.com}

\author[1]{\fnm{Kit} \sur{Windows-Yule}}\email{c.r.windows-yule@bham.ac.uk}

\affil*[1]{\orgdiv{School of Chemical Engineering}, \orgname{University of Birmingham}, \orgaddress{\city{Birmingham}, \country{United Kingdom}}}

\affil[2]{\orgname{Johnson Matthey plc}, \orgaddress{\city{London}, \country{United Kingdom}}}


\abstract{Designing and optimising the geometry of industrial process equipment remains slow and still largely ad hoc: engineers make small tweaks to one standard shape at a time, build prototypes, and hope for gains. We introduce HARPPP, an autonomous design loop that couples a compact, programmable geometry model to power-controlled CFD and evolutionary search. The geometry model is a single mathematical description that reproduces every standard impeller as a special case while spanning an unlimited set of manufacturable shapes. Working with Johnson Matthey on an industrial vessel, HARPPP explored a 23-parameter impeller–baffle space at constant power (3024 W), executing 3,000 simulation cycles in 15 days. The search uncovered multiple design families that outperform a Rushton/4-baffle baseline in both mixing intensity and uniformity, including twisted-plate impellers and pitched/curved baffles (intensity +18-78\%; uniformity CoV -16-64\%). A clear intensity–uniformity Pareto frontier emerged, enabling application-specific choices. Because HARPPP treats the simulator as the objective, it generalises to other equipment wherever credible physics models exist.}




\maketitle

\section*{Introduction}\label{intro}

\begin{figure*}[htbp]
\centering
\includegraphics[width=1.0\textwidth]{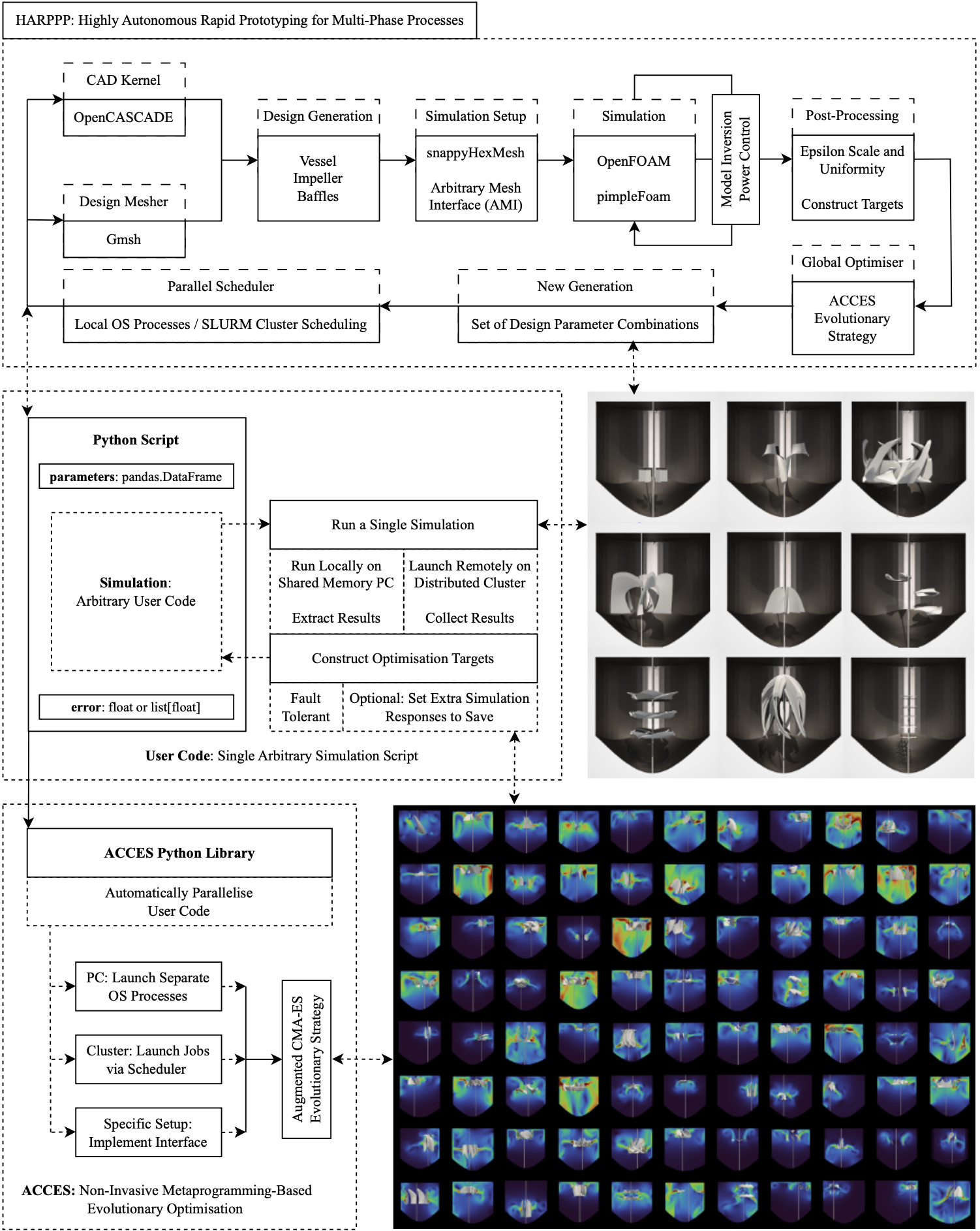}
\caption{\textbf{Architecture of HARPPP: Highly Autonomous Rapid Prototyping for Multi-Phase Processes.} 
HARPPP integrates parametric geometry generation, automated meshing, transient CFD simulation with model-inversion power control, and global optimisation in a closed-loop workflow. 
Design candidates are created through a programmed CAD kernel (forming a mathematical description of a geometry) and meshing tools before being simulated in OpenFOAM. 
Power control ensures fair comparison between geometries by maintaining constant energy input per volume. 
Post-processing evaluates turbulent dissipation and its uniformity, which are combined into scalar optimisation targets. 
An augmented CMA-ES evolutionary strategy (ACCES) generates new candidate geometries, with the entire cycle executed in parallel on local or cluster resources. 
Panels on the right illustrate representative impeller geometries and corresponding velocity fields explored during optimisation, showing their design diversity.}
\label{fig:intro}
\end{figure*}

Stirred tanks are among the most widely deployed unit operations in the process industries, spanning chemicals, pharmaceuticals, biotechnology and energy systems.\cite{paul2004handbook} Their ubiquity belies the difficulty of achieving application-specific flow characteristics: even small geometric changes--in impeller profile, placement, or baffle topology--can materially alter dissipation, circulation and shear, with consequences for micromixing, heat/mass transfer and reaction selectivity \cite{paul2004handbook,nienow1997mixing}. As a result, industrial design practice has historically relied on trial-and-error prototyping and pilot-scale trials, a path that is slow and capital-intensive. Design and scale-up of mixing-intensive unit operations remain empirically challenging and time-consuming, often requiring multi-stage experimentation and pilot testing with substantial engineering cost and schedule impact.

Computational fluid dynamics (CFD) offers a principled route to performance prediction and has seen extensive adoption in mixing research and engineering practice \cite{ranade2002cfd}. However, CFD has rarely been coupled to a \emph{fully autonomous} design loop at industrial scale, because each evaluation entails geometry generation, meshing, transient simulation and post-processing under realistic operating constraints. To date, studies typically restrict themselves to relatively simple geometric transformations (e.g., diameter, pitch, spacing -- for a given, fixed starting design), with complex topology changes largely out of scope \cite{heidari2025cfd,savage2024machine}. Consequently, optimisation campaigns in practice tend to explore narrow, human-curated families of geometries and terminate after tens to a few hundreds of designs -- insufficient for mapping the diversity of high-performing solutions in a high-dimensional parameter space, and overcome engineering bias from historical best practices.

In parallel, autonomous experimentation has transformed materials and catalysis, where self-driving laboratories and robotic chemists have demonstrated closed-loop exploration of complex spaces with dramatic throughput gains \cite{burger2020mobile,macleod2020self,hase2019next}. These systems leverage algorithmic search to surpass human sampling biases, uncovering non-intuitive regions of performance. Bringing the same philosophy to \emph{geometric} design in chemical engineering requires an end-to-end engine that (i) expresses broad design intent via a compact, manufacturable parameterisation; (ii) executes robust, power-constrained CFD evaluations at scale; and (iii) adapts sampling distributions to discover \emph{families} of high-performing solutions rather than collapsing to a single point optimum.

Crucially, an experimental analogue of this closed loop would be prohibitive. Fabricating and installing thousands of impeller/baffle variants demands workshop time, procurement, and vessel downtime; running each under matched power with sufficient steady-state sampling consumes substantial utilities and operator time while generating waste. More fundamentally, the key observables we optimise -- field-resolved turbulent dissipation $\varepsilon$ and its spatial statistics -- are not directly measurable in opaque, industrial vessels; high-fidelity PIV/LDV requires transparent, index-matched models and sub-Kolmogorov resolution, and even then yields sparse, near-wall-biased data \cite{baldi2003direct, escudie2003experimental}. By contrast, the CFD-in-the-loop workflow delivers volume-resolved $\varepsilon$ for \emph{every} candidate at constant power, safely screens extreme or unstable geometries before any hardware is cut, and enables scale-free parallel exploration (e.g.\ the 3{,}000 design–simulation cycles reported here). In short, simulation makes the “self-driving” paradigm feasible for geometric design at industrial scale -- experiment alone does not.

Here we introduce \textsc{HARPPP} -- Highly Autonomous Rapid Prototyping for Multi-Phase Processes -- a fully automated pipeline that couples a parametric CAD kernel to transient, power-controlled CFD within an evolutionary optimisation loop (Figure \ref{fig:intro}). Our optimiser builds on the covariance-matrix adaptation evolution strategy (CMA-ES), whose rank-based, affine-invariant updates and self-adaptation of the covariance and step size are well-suited to expensive, noisy black-box objectives \cite{nicusan2025acces,hansen2001derandomized}. Our novel framework builds upon CMA-ES to facilitate in-loop parallel sampling, phenotype normalisation and native archiving, enabling us to treat the entire simulator as the objective while preserving exploration across heterogeneous geometric descriptors.

To ensure genuine industrial relevance, we conduct the case study with Johnson Matthey (JM) using a JM-owned benchmark vessel and JM’s operating constraints, evaluating all candidates at matched power (3024\,W) for fair comparison. This anchors the work in real plant practice rather than an academic surrogate.

Using \textsc{HARPPP}, we executed 3{,}000 autonomous design-simulation cycles (1{,}500 impellers and 1{,}500 baffles) in 15 days. The search revealed multiple, morphologically distinct families that significantly exceeded a baseline Rushton/baffled configuration on both mixing intensity (volume-averaged $\mathrm{mean}(\varepsilon)$) and uniformity (coefficient of variation, $\mathrm{CoV}(\varepsilon)$), including simple, easily manufacturable twisted-plate impellers and material-efficient, pitched/curved baffles that reduce $\mathrm{CoV}(\varepsilon)$ without relying on four full-height vertical plates. Beyond individual optima, the archive delineates a clear Pareto frontier between intensity and uniformity, enabling application-specific selection (e.g.\ favouring higher local dissipation for micromixing-limited processes, or lower $\mathrm{CoV}(\varepsilon)$ where homogeneous energy input is paramount). 

Our results demonstrate that autonomous, simulator-in-the-loop optimisation can look beyond traditional design principles, efficiently charting high-performing regions in a 23-dimensional geometric space under realistic operating constraints. The framework is general: it requires no problem-specific surrogates or hand-crafted features and can be ported to other reactor internals or process equipment wherever a simulation engine (e.g. CFD, DEM, FEM) provides a credible physics model.

\section*{Results}\label{results}

We evaluated HARPPP on an industrial-scale stirred tank reactor design supplied by Johnson Matthey, chosen to avoid intellectual-property-sensitive geometries while providing realistic constraints for optimisation. The baseline configuration consisted of a rounded-bottom cylindrical vessel equipped with a centred six-blade Rushton turbine and four baffles (full details included in Table \ref{tab:baseline}).

To ensure fair comparison between candidate designs, all subsequent impeller and baffle geometries were simulated at a dynamically adjusted rotation rate such that the net power input matched the JM-supplied 3024\,W baseline value. This avoids degenerate cases where unconventional geometries could appear artificially effective at fixed speed while requiring significantly greater power inputs. Performance improvements reported here are therefore attributable solely to geometry, rather than disparities in operating conditions.

\begin{table*}[htbp]
\centering
\caption{Baseline stirred tank configuration provided by Johnson Matthey and used as the reference case for optimisation.}
\label{tab:baseline}
\begin{tabular}{l c c}
\hline
\textbf{Parameter} & \textbf{Symbol} & \textbf{Value} \\
\hline
Vessel diameter & $T$ & 2.00 m \\
Liquid height (from bottom of dished base) & $H$ & 2.00 m \\
Liquid height (from bottom tan line) & $S$ & 1.40 m \\
Rushton turbine diameter (blade tips) & $D$ & 0.67 m \\
Rushton turbine blade height & $W$ & 0.13 m \\
Rushton turbine blade length & $L$ & 0.17 m \\
Rushton turbine blade thickness & $X$ & 0.01 m \\
Rushton turbine disc diameter & $D_d$ & 0.40 m \\
Dished end height & $Z$ & 0.60 m \\
Number of baffles & -- & 4 \\
Baffle width & $W_b$ & 0.17 m \\
Baffle clearance from wall & $W_s$ & 0.03 m \\
Fluid density & $\rho$ & 997 kg\,m$^{-3}$ \\
Fluid viscosity & $\mu$ & $8.91\times 10^{-4}$ Pa\,s \\
Approximate volume & -- & 4.40 m$^{3}$ \\
Baseline impeller speed & $N$ & 100 - Variable \\
Baseline Reynolds number & $Re$ & $8.3\times10^{5}$ - Variable \\
Target power input (all designs) & $P$ & 3024 W - Constant \\
\hline
\end{tabular}
\end{table*}

\subsection*{Simulation Validation}\label{sec:validation}

\begin{figure*}[h]
\centering
\includegraphics[width=1.0\textwidth]{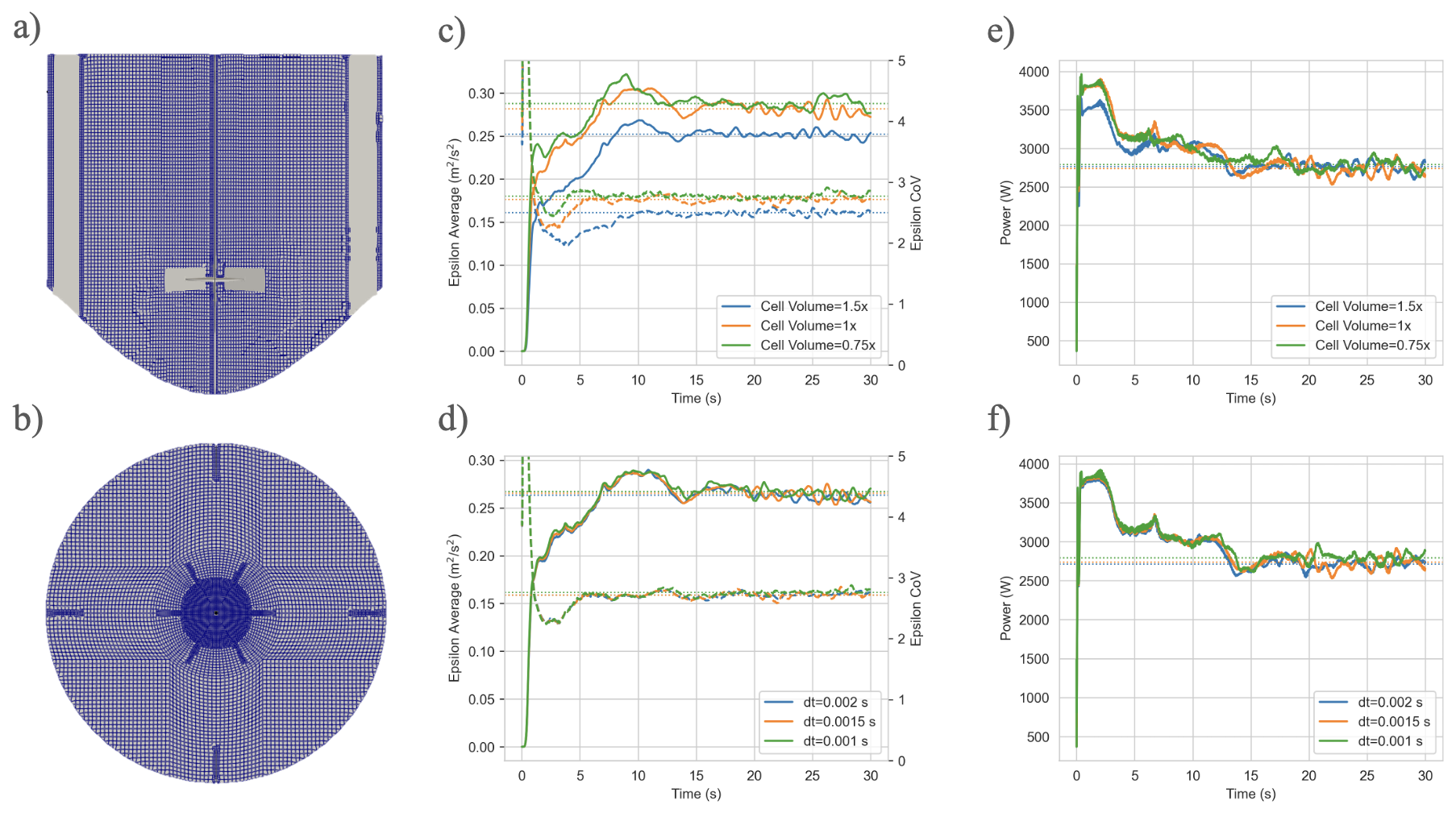}
\caption{\textbf{Validation of the CFD framework for stirred tank simulations.} 
(a,b) Representative cross-sectional and top views of the computational mesh used to resolve the baffled stirred tank geometry for the base case. 
(c,d) Mesh-independence study showing the time evolution of turbulent dissipation rate ($\varepsilon$, solid lines, left axis) and its coefficient of variation (CoV, dashed lines, right axis), together with the corresponding power draw, for coarse ($4.3\times 10^5$ cells), medium ($6.2\times 10^5$ cells), and fine ($8.5\times 10^5$ cells) meshes. 
(e,f) Timestep-independence study showing $\varepsilon$, CoV, and power draw for timesteps $\Delta t=0.001$-$0.002$ s. 
Both $\varepsilon$ and power stabilise with negligible variation across mesh and timestep refinements, confirming numerical convergence and validating the framework for subsequent optimisation studies.}
\label{fig:validation}
\end{figure*}

To ensure that the optimisation results were not biased by numerical artefacts, we first performed mesh and timestep independence studies and validated the power-control scheme (Fig.~\ref{fig:validation}).

\subsubsection*{Mesh Independence}

Three meshes were tested, with $4.3\times10^5$, $6.2\times10^5$, and $8.5\times10^5$ cells. 
Volume-averaged turbulent dissipation rate ($\varepsilon$), its coefficient of variation (CoV), and impeller power draw were compared over the final 15\,s of each simulation. 
The medium mesh reproduced the fine-mesh averages within 2.1\% for $\varepsilon$, 1.8\% for CoV, and 1.9\% for power, while requiring substantially less computational time. 
Errors for the coarse mesh were higher, reaching 12.4\% in $\varepsilon$ and 9.8\% in CoV. 
We therefore adopted the medium mesh ($6.2\times10^5$ cells) as a reliable balance between accuracy and efficiency.  
These resolutions are consistent with prior CFD studies of stirred tanks of comparable size, which typically employed meshes of $0.5$-$2.0\times10^6$ cells \cite{singh2011assessment}.

\subsubsection*{Timestep Independence}

Timesteps of $\Delta t = 0.001$-$0.002$,s were tested, corresponding to angular displacements of $0.6$-$1.2^\circ$ per impeller revolution at 100\,rpm.
Average $\varepsilon$, CoV, and power draw all agreed within 2.8\% across this range, with no systematic trends. 
We selected $\Delta t=0.0015$\,s to ensure numerical stability for HARPPP-generated designs.

\subsubsection*{Power Control Validation}

Across all mesh and timestep refinements, the simulation-in-the-loop controller -- detailed in Methods -- successfully adjusted the impeller rotation rate to maintain the target power input of 3024\,W (see Figure \ref{fig:fields} for the continuous adjustment curve of RPM vs. Power response).

\subsubsection*{Baseline Case}

\begin{figure*}[htbp]
\centering
\includegraphics[width=1.0\textwidth]{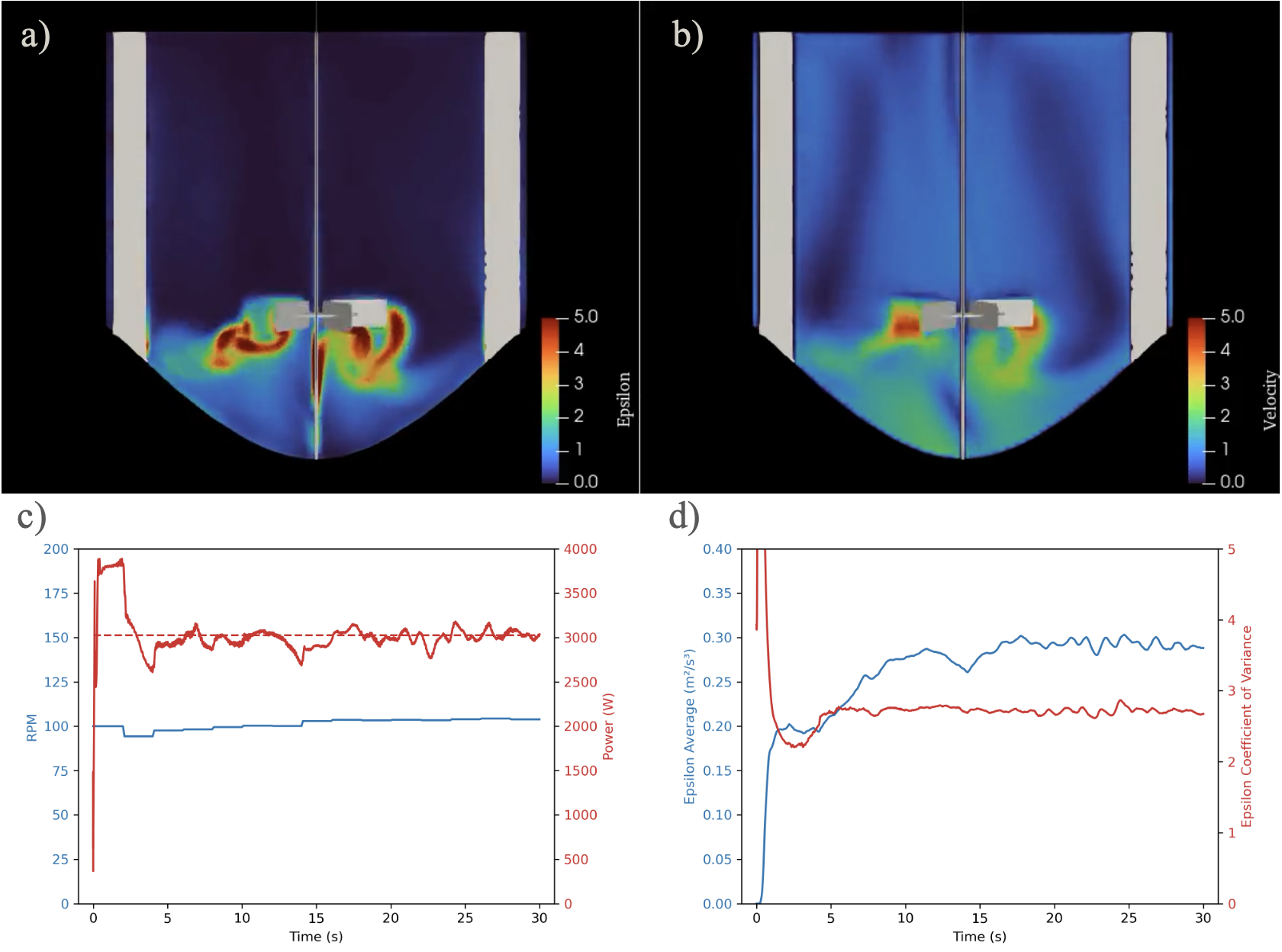}
\caption{\textbf{Baseline Rushton turbine performance at constant power input.} 
(a) Turbulent dissipation rate ($\varepsilon$) and (b) velocity magnitude fields in the baffled stirred tank at steady state, showing strong localised dissipation in the impeller discharge stream. 
(c) Dynamic power-control behaviour: the impeller rotation rate (blue) is continuously adjusted to maintain a target input power of 3024\,W (red). 
(d) Evolution of mean $\varepsilon$ (blue) and its coefficient of variation (red), showing stabilisation within 10\,s. 
These results establish the baseline hydrodynamics against which optimised geometries are compared.}
\label{fig:fields}
\end{figure*}

Having established numerical convergence, we next characterised the baseline Rushton configuration at the target power input of 3024 W. The dynamic controller adjusted the impeller speed during start-up to maintain this power level (Fig. \ref{fig:fields}c), after which both the mean turbulent dissipation and its coefficient of variation stabilised (Fig. \ref{fig:fields}d). The corresponding spatial distributions of dissipation and velocity (Fig. \ref{fig:fields}a,b) show strongly localised energy input in the impeller discharge stream, consistent with the flow physics of Rushton turbines \cite{kratvena2004study, lee1998turbulence}. These baseline results provide the hydrodynamic reference against which optimised geometries were evaluated.

\subsection*{Optimisation Problem}

\subsubsection*{Geometric Parametrisation}

Our CAD kernel is not a catalogue of shapes but a \emph{compact mathematical language} for stirred-tank internals. Canonical industrial impellers are recovered as \emph{low-dimensional submanifolds} (parameter constraints) of the same general descriptor: a Rushton turbine corresponds to zero twist/helix with rectangular blades on a disc (\texttt{twist}=\texttt{helix}=0, constant chord/thickness, \texttt{num\_blades}=6); pitched-blade turbines impose a constant pitch via \texttt{turn} angles and zero \texttt{helix}; axial propellers arise from nonzero \texttt{helix} and cambered profiles; high-shear devices use small chord, large tip speed, and high curvature; ribbon/anchor types emerge from large span, high aspect ratio and continuous wrap-around with axial repeats. In other words, “Rushton”, “PBT”, “propeller”, “ribbon” etc.\ are \emph{special cases} of a single, unified parameterisation; relaxing those equalities yields a \emph{continuum} (effectively infinite) of novel, manufacturable geometries generated by smooth deformations (twist, helix, joint/lean/turn, profile, repeats). This generality lets the optimiser search \emph{beyond} human-curated design families while still reproducing any standard design on demand (Figure \ref{fig:profiles}). Full details of the blade profile parametrisation are provided in Methods.

\subsubsection*{Optimisation Objectives}

\textbf{Choosing targets.} In stirred tanks, many performance metrics are controlled by the local turbulent kinetic–energy dissipation rate, \(\varepsilon\). The Kolmogorov time and length scales,
\[
\tau_\eta \sim \left(\nu/\varepsilon\right)^{1/2}, \qquad
\eta \sim \left(\nu^{3}/\varepsilon\right)^{1/4},
\]
set micromixing rates and the finescale at which scalar gradients are dissipated. Larger \(\mathrm{mean}(\varepsilon)\) (at \emph{fixed power}) implies shorter micromixing times and higher scalar dissipation, which in practice improves selectivity in mixing-sensitive competitive reactions, increases gas–liquid mass-transfer coefficients (\(k_La\) scales positively with \(\varepsilon\) in standard correlations), and drives smaller droplet and crystal sizes (e.g., Hinze-type break-up scalings). Equally important, a low spatial variability of \(\varepsilon\) (low \(\mathrm{CoV}(\varepsilon)\)) suppresses dead zones and hot spots, reducing batch-to-batch scatter, avoiding over-shear (relevant for cells/enzymes), and yielding more uniform product attributes. Operating every candidate at the \emph{same} power therefore lets geometry that converts input power into useful finescale mixing (rather than bulk recirculation) rise to the top.%
\footnote{See, e.g., standard treatments in \cite{paul2004handbook,ranade2002cfd} for links between \(\varepsilon\), micromixing, mass transfer and dispersion.}

\medskip
\noindent\textbf{Formulation.} We evaluated several scalar targets. Maximising \(\mathrm{mean}(\varepsilon)/\sigma(\varepsilon)\) (the inverse coefficient of variation, \(\mathrm{CoV}^{-1}\)) rewards both intensity and homogeneity but can be numerically brittle as \(\sigma\!\to\!0\). Geometric and harmonic means were also tested; the former under-penalises inhomogeneity, the latter can be dominated by extreme lows. We therefore adopted two complementary objectives and a simple scalarisation:
\begin{align}\label{eq:targets}
    F_1 &= -\mathrm{mean}(\varepsilon),\\
    F_2 &= \frac{1}{\mathrm{CoV}(\varepsilon)},\\
    \Leftrightarrow\quad F_1 \cdot F_2 &= \frac{-\mathrm{mean}^2(\varepsilon)}{\sigma(\varepsilon)}.
\end{align}
Minimising \(F_1\!\cdot\!F_2\) simultaneously drives higher \(\mathrm{mean}(\varepsilon)\) and lower \(\mathrm{CoV}(\varepsilon)\). Although this scalarisation does not enumerate the full Pareto front (which would require far more evaluations), it reliably steers search toward design \emph{families} that deliver both stronger and more uniform finescale mixing. As shown later, an intensity–uniformity trade-off still emerges naturally.

\medskip
\noindent\textbf{Validity filter.} Only simulations that remained numerically stable for the full 30\,s horizon were retained; failed runs were marked NaN and excluded from selection to prevent spurious optima.

\subsubsection*{Optimiser}

We employed a large-scale evolutionary strategy that we developed for simulation-in-the-loop optimisation and design discovery. This is part of the HARPPP workflow: for each candidate, the optimiser emits a parameter set, the metaprogramming layer materialises a run (simulation script → parameter files → geometry generation → meshing → simulation), and the post-processing stage saves objective values which are collected by the head node to drive the next evolutionary update (see Figure \ref{fig:intro}).

To make the search well-conditioned across heterogeneous kernel variables, all parameters were affine-scaled to a common phenotype space such that, at initialisation, each dimension had comparable scale, with unit variance. The initial search distribution was set isotropic with a standard deviation of 0.4 in this space, providing broad, near-uniform coverage of the admissible design ranges while remaining compatible with box constraints (handled by resampling/clipping). The mean and covariance of the search distribution were then adapted generation-by-generation using rank-based selection and cumulative step-size control, as in CMA-ES \cite{hansen2003reducing}.

Objective evaluation proceeded in parallel batches, in this case set to a family size of \(\lambda = 50\) designs per generation. Simulations that did not remain stable for the full 30\,s horizon were marked invalid and excluded from selection, preventing numerical failures from biasing adaptation. For selection we used the scalarised score formed from the two objectives $F_1, F_2$, while retaining the individual components for analysis.

In this work the evolutionary strategy was used primarily for \emph{guided parameter exploration and rapid identification of well-performing design families}, rather than for converging to a single point optimum. We therefore emphasised broad coverage in early generations and terminated on a fixed evaluation budget or plateau in the scalarised objective; for these particular problems we stopped after 30 generations, as the best design families were identified in earlier epochs, with no significantly different design families emerging.

\subsection*{Impeller optimisation}

\begin{figure*}[htbp]
\centering
\includegraphics[width=1.0\textwidth]{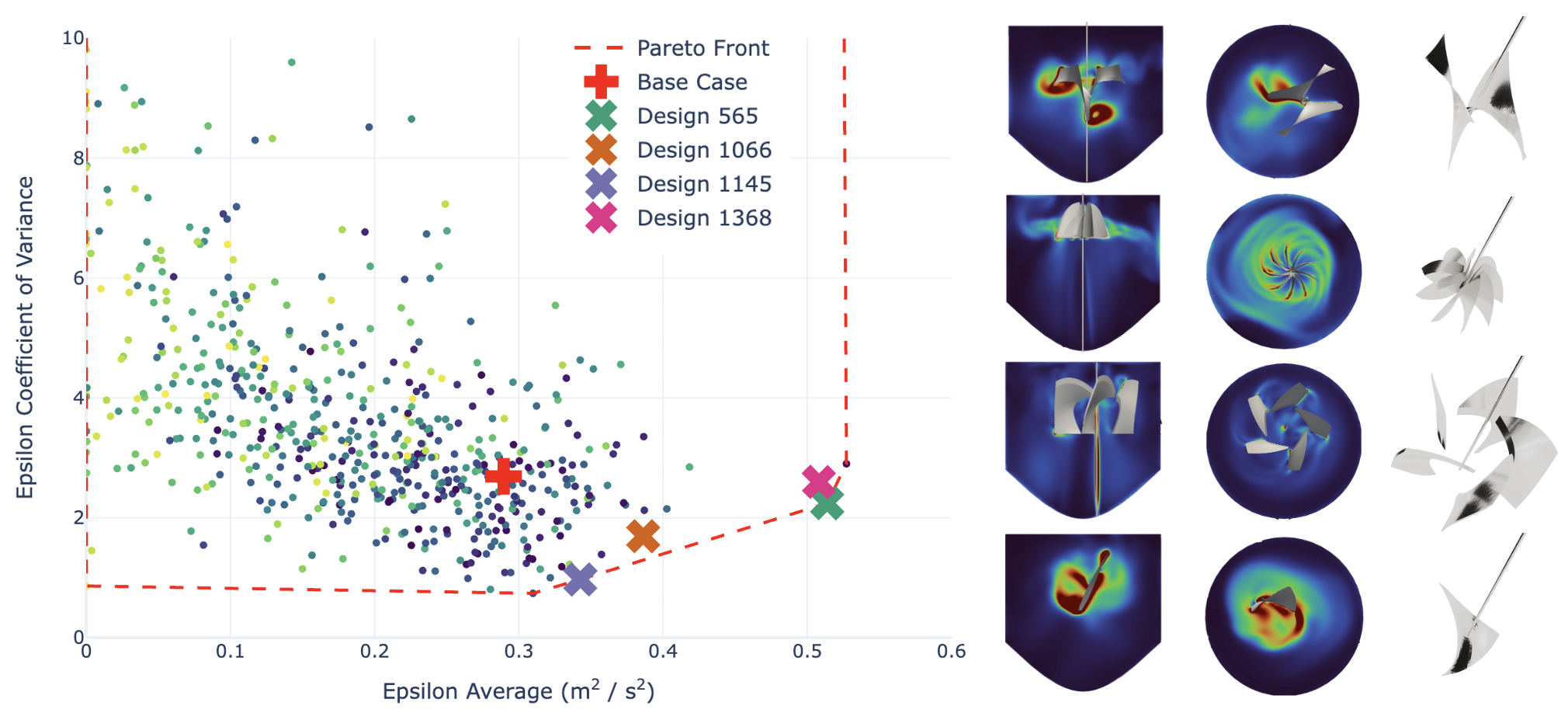}
\caption{\textbf{Impeller optimisation landscape and representative families.}
\emph{Left:} Scatter of all evaluated designs in the 23-parameter impeller space,
plotted by volume-averaged turbulent dissipation $\mathrm{mean}(\varepsilon)$
($\mathrm{m}^2\,\mathrm{s}^{-3}$) versus its coefficient of variation
$\mathrm{CoV}(\varepsilon)$ (dimensionless). Better performance lies to the right
(higher dissipation) and downward (greater uniformity). Later generations are shown as darker points. The dashed curve marks the
non-dominated (Pareto) front. The baseline Rushton configuration is shown as a red “+”,
and four exemplars (Designs 565, 1066, 1145, 1368; crosses) are highlighted.
\emph{Right:} Corresponding geometries and instantaneous $\varepsilon$ fields
(cross-section and top view; colour bar 0--5\,$\mathrm{m}^2\,\mathrm{s}^{-3}$),
illustrating distinct design families spanning the intensity--uniformity trade-off.}
\label{fig:impeller_opt}
\end{figure*}

Figure~\ref{fig:impeller_opt} summarises the optimisation over the
23-dimensional impeller parameterisation (see Methods for the full list of descriptors).
Although a scalarised objective was used in-loop, we report the constituent axes
$\mathrm{mean}(\varepsilon)$ and $\mathrm{CoV}(\varepsilon)$ to expose the trade-off.

In total, $\lambda=50$ designs per generation were evaluated over 30 generations
($N=1{,}500$ impellers) at fixed input power (3024\,W). Simulations that did not reach
30\,s of physical time were censored (objective set to NaN) and excluded from selection.

The baseline Rushton case achieved
$\mathrm{mean}(\varepsilon)=0.289$ and $\mathrm{CoV}(\varepsilon)=2.691$.
The four exemplars improved on both axes:

\begin{itemize}
\item \textbf{Design 565}: $\mathrm{mean}(\varepsilon)=0.514$,
$\mathrm{CoV}(\varepsilon)=2.244$ \;(\,+77.6\% intensity, $-16.6\%$ CoV vs. baseline\,);
a high-intensity solution with simple topology (two twisted plates).
\item \textbf{Design 1368}: $\mathrm{mean}(\varepsilon)=0.508$,
$\mathrm{CoV}(\varepsilon)=2.598$ \;(\,+75.5\% intensity, $-3.5\%$ CoV\,);
also high-intensity, manufacturable as a single twisted plate.
\item \textbf{Design 1066}: $\mathrm{mean}(\varepsilon)=0.386$,
$\mathrm{CoV}(\varepsilon)=1.693$ \;(\,+33.4\% intensity, $-37.1\%$ CoV\,);
closest to a conventional vertical-blade family (upper edges shaved), placed higher in the tank.
\item \textbf{Design 1145}: $\mathrm{mean}(\varepsilon)=0.343$,
$\mathrm{CoV}(\varepsilon)=0.971$ \;(\,+18.5\% intensity, $-63.9\%$ CoV\,);
the most uniform of the set.
\end{itemize}

These results delineate a clear Pareto frontier between mixing intensity and spatial
uniformity. Notably, substantial gains over the baseline do not require intricate
geometries: simple twisted-plate families (565, 1368) attain the highest intensities,
whereas 1145 prioritises uniformity. This enables application-specific selection--e.g., for micromixing-limited reactions,
prioritising \emph{higher local turbulent dissipation} (and thus shorter micromixing times),
whereas processes requiring homogeneous energy input prioritise \emph{lower}
$\mathrm{CoV}(\varepsilon)$.

\subsubsection*{Regularities among near-front designs}

Across the non-dominated impellers at 3024\,W (23-D search), two striking regularities emerged:
(i) \emph{no} near-front design used axial repeats (\texttt{repeat\_number}=1 throughout), indicating no benefit from stacking multiple impeller planes under a fixed power budget; and
(ii) \emph{all} near-front designs employed deliberate shaft eccentricity,
$r_{\mathrm{ecc}}=\sqrt{x_{\mathrm{place}}^{2}+y_{\mathrm{place}}^{2}}>0$, which broadened the footprint of elevated $\varepsilon$ and consistently reduced $\mathrm{CoV}(\varepsilon)$ relative to centred shafts.
This directly challenges standard practice for tall vessels (multiple impeller stages) and the near-universal assumption of centred mounting for single-stage mixing \cite{paul2004handbook,nienow1997mixing,ranade2002cfd}.

\medskip
\noindent\textbf{Symmetry breaking redistributes dissipation.}
At constant power, stacking divides $P$ across planes, lowering local blade tip speeds and concentrating energy into repeated, mirror-symmetric discharge structures that do little to reduce global $\mathrm{CoV}(\varepsilon)$. By contrast, a single plane with \emph{eccentric placement} breaks azimuthal symmetry: the high-$\varepsilon$ jet precesses relative to the tank/baffles, engages the wall more uniformly, and disrupts coherent recirculation cells. The result is a wider, more evenly distributed dissipation field at the same net $P$ (Fig.~\ref{fig:impeller_opt}), i.e.\ higher useful finescale mixing per watt and lower dead-zone risk -- without resorting to additional stages.

\medskip
\noindent\textbf{Industrial significance.}
This suggests a simpler path to uniformity at scale: \emph{one} effective impeller plane, deliberately off-centre, instead of multi-stage stacks. That implies fewer components, easier retrofits, simpler CIP/inspection, and lower CAPEX/OPEX, while delivering better intensity–uniformity trade-offs.

\subsection*{Baffle Optimisation}

\begin{figure*}[htbp]
\centering
\includegraphics[width=1.0\textwidth]{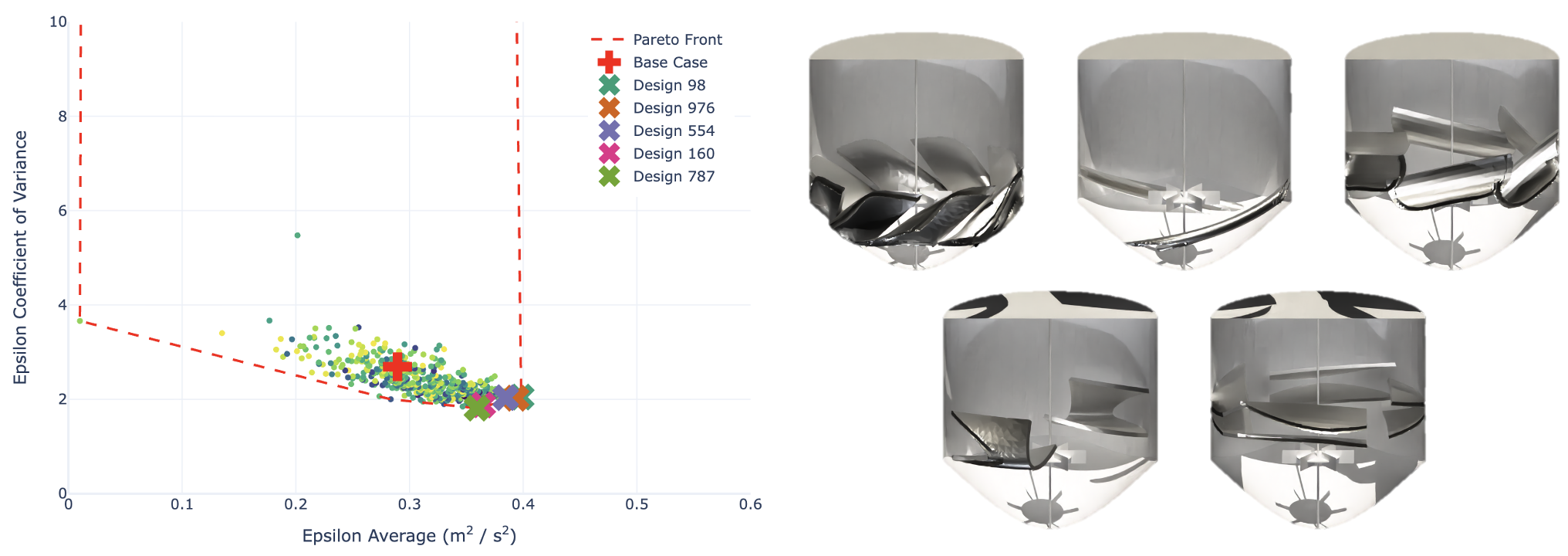}
\caption{\textbf{Baffle optimisation landscape and representative families.}
\emph{Left:} Scatter of all evaluated baffle designs, shown in the plane of
volume-averaged dissipation $\mathrm{mean}(\varepsilon)$ versus
$\mathrm{CoV}(\varepsilon)$. Higher intensity lies to the right; greater
uniformity downward. Later generations are shown in darker colours. The red dashed curve denotes the non-dominated set.
The baseline configuration is marked with a red “+”; five exemplars
(designs 98, 976, 554, 160, 787) are highlighted. 
\emph{Right:} Corresponding baffle geometries (empty tank shown for clarity),
illustrating the morphological diversity that attains near-front performance.}
\label{fig:baffles_opt}
\end{figure*}

Figure~\ref{fig:baffles_opt} summarises the baffle search using the same
two axes as for the impeller.
The baseline achieved $\mathrm{mean}(\varepsilon)=0.289$\,m$^2$\,s$^{-3}$ and
$\mathrm{CoV}(\varepsilon)=2.691$.
The five highlighted designs all improved upon the baseline on both axes,
with increases in $\mathrm{mean}(\varepsilon)$ of $+24$-$+38\%$ and reductions in
$\mathrm{CoV}(\varepsilon)$ of $-24$-$-33\%$:

\begin{itemize}
\item \textbf{Design 98:} $\mathrm{mean}(\varepsilon)=0.398$, 
$\mathrm{CoV}(\varepsilon)=2.047$ \;(+$37.6\%$, $-23.9\%$ vs.\ baseline).
\item \textbf{Design 976:} $\mathrm{mean}(\varepsilon)=0.394$, 
$\mathrm{CoV}(\varepsilon)=2.023$ \;(+$36.1\%$, $-24.8\%$).
\item \textbf{Design 554:} $\mathrm{mean}(\varepsilon)=0.385$, 
$\mathrm{CoV}(\varepsilon)=2.035$ \;(+$32.9\%$, $-24.4\%$).
\item \textbf{Design 160:} $\mathrm{mean}(\varepsilon)=0.364$, 
$\mathrm{CoV}(\varepsilon)=1.881$ \;(+$25.9\%$, $-30.1\%$).
\item \textbf{Design 787:} $\mathrm{mean}(\varepsilon)=0.359$, 
$\mathrm{CoV}(\varepsilon)=1.813$ \;(+$24.2\%$, $-32.6\%$).
\end{itemize}

Across near-front solutions the gains are driven primarily by improved
uniformity: the best designs cluster around
$\mathrm{mean}(\varepsilon)\approx 0.36$-$0.40$\,m$^2$\,s$^{-3}$ while
$\mathrm{CoV}(\varepsilon)$ drops to $\approx 1.8$-$2.05$.
This indicates that, within the explored parameter space, baffles chiefly
re-distribute dissipation rather than pushing intensity to the very highest
levels; nevertheless, several families deliver simultaneous increases in
$\mathrm{mean}(\varepsilon)$ and marked reductions in $\mathrm{CoV}(\varepsilon)$
relative to the baseline. 
The five exemplars in Fig.~\ref{fig:baffles_opt} illustrate that such
performance can be achieved by diverse shapes (curved and pitched segments,
truncated arcs), supporting multiple manufacturable routes to improved
uniformity.

\subsubsection*{Morphology and material efficiency.}
A notable regularity among the near-front baffle solutions is structural economy: many are composed of short, curved or pitched segments rather than four full-height vertical plates, implying substantially lower solid volume (and, by extension, material usage) than the canonical design. Equally, none of the top designs are purely vertical; instead they favour elements with pronounced pitch or curvature that redirect the flow quasi-horizontally rather than relying on strong normal blockage and wake break-up. This morphology is consistent with the observed reductions in $\mathrm{CoV}(\varepsilon)$, suggesting that gentle momentum redirection and distributed shear--rather than hard impingement--provides a more uniform dissipation field at fixed power input.

\section*{Discussion}\label{discussion}

\subsection*{Key findings}
Using a constant power budget (3024\,W) to ensure that differences in turbulence dissipation are due to geometric effects and not changes in volumetric power input, HARPPP uncovered multiple impeller and baffle \emph{design families} that dominate the industry-standard baseline in both average dissipation and spatial uniformity. For impellers (Fig.~\ref{fig:impeller_opt}), twisted-plate morphologies (Designs 565, 1368) achieved the largest increases in $\mathrm{mean}(\varepsilon)$ (up to $\,+78\%$) while also reducing $\mathrm{CoV}(\varepsilon)$ relative to the Rushton reference; a second family (Design 1145) prioritised uniformity ($-64\%$ CoV) at only moderate intensity penalty compared to the other optimised designs. For baffles (Fig.~\ref{fig:baffles_opt}), multiple non-canonical shapes improved both axes simultaneously (up to $\,+38\%$ in $\mathrm{mean}(\varepsilon)$ and $-33\%$ in $\mathrm{CoV}(\varepsilon)$), despite some using substantially less material than four full-height plates.

\subsection*{Physical interpretation}
Two regularities recur among near-front impellers. First, \emph{deliberate eccentricity} ($r_{\mathrm{ecc}}>0$) is consistently favoured: breaking azimuthal symmetry spreads the high-shear footprint away from a single, coherent jet, lowering $\mathrm{CoV}(\varepsilon)$ without sacrificing intensity. Second, \emph{axial stacking} did not confer an advantage under a fixed power budget: repeats add swept volume and tip losses without proportionate gains in cross-tank transport, so the optimiser concentrates power into one effective plane.

Among baffles, the best designs are not purely vertical. Instead, short pitched or curved segments preferentially \emph{redirect} flow quasi-horizontally rather than relying on strong normal blockage and wake breakup. This gentler momentum turning creates more distributed shear, which aligns with the observed reduction in $\mathrm{CoV}(\varepsilon)$. An additional practical benefit is structural economy: many near-front baffles use far less solid volume than the canonical four-plate design, implying lower material and installation cost at comparable or improved hydrodynamic performance.

\subsection*{Exploration over single-point optimisation}
The evolutionary strategy was used deliberately as a \emph{design discovery} tool rather than to force single-point convergence. Rank-based, invariant ACCES optimisation with CMA-ES updates adapt the sampling ellipsoid to the sensitive directions of the response, revealing elongated, high-performing manifolds in parameter space (design families) while maintaining diversity. Archiving every evaluated design then enables post hoc reconstruction of the dissipation-uniformity Pareto frontier, even though a scalarised score guided the in-loop selection. This workflow yields actionable choices for different process aims (e.g., higher local dissipation for reactions limited by sub-Kolmogorov mixing versus lower $\mathrm{CoV}(\varepsilon)$ where homogeneous energy input is paramount) without requiring a priori weights for a multiobjective solver.

\subsection*{Implications for deployment}
Several of the highest-intensity impellers are manufacturable as simple twisted plates (one or two elements), suggesting that substantial performance gains over Rushton turbines do not require complex geometries. Likewise, near-front baffle sets favour compact, pitched/curved segments; these could retrofit existing tanks with minimal mechanical changes and reduced material, potentially simplifying CIP (clean-in-place) and maintenance.

\paragraph{Human-in-the-loop selection, by design.}
HARPPP \emph{does not} auto-deploy a single “best” geometry. Instead, it surfaces \emph{families} of high-performing, manufacturable options and a transparent archive of every evaluation, exposing the intensity–uniformity Pareto trade-off and the geometric choices that produce it. This deliberately keeps domain experts “in the loop”: process engineers can select designs that align with plant-specific goals (e.g., micromixing limits vs.\ bulk homogeneity), constraints (materials, CIP, fouling, retrofittability), and governance (safety cases, GMP/GxP). Ethically, this prevents hidden optimiser preferences from becoming de facto design policy and preserves accountability; practically, it accelerates decision-making with auditable evidence while respecting expert judgement and local operating knowledge.

\paragraph{From case study to platform.}
The same loop used here generalises along four axes: \emph{physics} (CFD $\rightarrow$ DEM/FEM/LES/reacting flows), \emph{normalisation} (constant power $\rightarrow$ constant $\Delta p$, throughput, energy, or material), \emph{constraints} (collision, minimum radii, cleanability, GMP traceability), and \emph{search} (single $\rightarrow$ multiobjective/robust/multifidelity with coarse-to-fine refinement). Because every evaluation is archived with geometry, operating point, and metrics, HARPPP provides governance-ready provenance and keeps experts in the decision loop, turning the output into ranked, manufacturable families matched to process aims rather than a black-box “best” design.

\subsection*{Concluding remark}
HARPPP is a general, simulator-in-the-loop platform for geometric design. Its three primitives—(i) a programmable, manufacturability-aware geometry kernel; (ii) resource-normalised evaluations inside any credible physics engine (e.g., power/pressure/throughput control in CFD, particle count/throughput in DEM, stress/weight in FEM); and (iii) distribution-based evolutionary search with native parallel orchestration and full archiving -- compose into a reusable workflow for industrial geometric design optimisation. Swapping the kernel and evaluator makes it possible to target heat-exchanger inserts, static mixers, cyclones, granular mixers, blade mills, atomisers, battery flow plates, and beyond, while optimising objectives such as yield, pressure drop, shear history, residence-time variance, stress, cost or carbon footprint (single-, multiobjective, or robust). Critically, HARPPP discovers \emph{families} of manufacturable designs rather than a single point, enabling human-in-the-loop selection under plant, regulatory, and sustainability constraints. In short, HARPPP turns equipment design from ad hoc iteration into a scalable, auditable, and defensible process that is portable across unit operations and physics.

\section*{Methods}\label{methods}

\subsection*{Geometric Parametrisation}

Our parametrisation begins with a blade profile that is scaled to a given chord length and thickness, then displaced radially from the impeller root, extruded along the rotation axis and optionally deformed by twist (rotation about the chord) and helix (rotation about the impeller root). Blade elements are further articulated at their root by three sequential rotations: joint, lean, and turn, about given axes along their characteristic lengths. The full impeller is assembled by repeating the blade radially and optionally stacking repeats axially with controlled spacing and angular bias. Baffles are generated using the same kernel, then implemented as stators in the simulation.

The goal of this parametrisation was to capture a very wide variety of blade shapes--including C- and S-curved profiles and multi-modal contours, as well as comparatively flat Rushton-type turbines--using only a few interpretable parameters. Classical airfoil parameterisations (e.g. IGP with 8 parameters \cite{lu2018improved} or PARSEC with 10 parameters \cite{derksen2010bezier}) are often too restrictive for stirred-tank geometries: they impose aerodynamic continuity and leading-edge constraints that are less relevant in mixing contexts. Our approach, by contrast, allows intentional “open” profiles and sharp curvatures, which HARPPP can exploit to influence local turbulent dissipation (examples given in Figure \ref{fig:profiles}).

\subsubsection*{Blades profile}

HARPPP generates blade sections from a compact, manufacturable curve family with
2--3 descriptor parameters that control effective camber/curvature and thickness, including non-monotonic variation.
Let $\boldsymbol{\theta}\in\Theta$ denote the low-dimensional descriptor for a blade side.
Internally, a directly-programmed CAD generator maps $\boldsymbol{\theta}$ to a watertight,
non-self-intersecting planar profile, aligns it to a chord, and produces an extrusion-ready surface:

\begin{align*}
\mathcal{G}:\ \boldsymbol{\theta}\ \mapsto\ \text{CAD profile} \;\;\Rightarrow\;\;\\
\text{watertight solid compatible with meshing}.
\end{align*}

\paragraph{Interface and guarantees.}
For any admissible $\boldsymbol{\theta}\in\Theta$ and target chord length $L$ and
thickness $S$, the generator returns a blade side with:
(i) chord-aligned normalisation to $[0,L]$,
(ii) bounded thickness $S$ with manufacturability-enforced minimum radii,
(iii) $C^0$ watertightness and non-self-intersection,
and (iv) determinism (identical input yields identical CAD). These guarantees ensure
meshing robustness and comparability across candidates.

\paragraph{Two-sided profiles and validity.}
Complete blade sections are formed by pairing two admissible descriptors
$\boldsymbol{\theta}^{(1)},\boldsymbol{\theta}^{(2)}$ on a common chord and thickness.
If an assembled section violates any geometric constraint (e.g.\ self-intersection or
local radius limits), HARPPP applies an internal, deterministic correction or rejects
the candidate prior to meshing.

\paragraph{Placement and higher-order deformations.}
Blade placement and articulation (e.g.\ joint/lean/turn, twist, helix) are applied to the
solid section after profile generation using transformations which can be bounded by manufacturability constraints.
These controls allow us to reproduce canonical industrial designs as low-dimensional
submanifolds and to explore continuous deformations beyond industry-standard designs, while
respecting mechanical and cleanability constraints.

\paragraph{Blade profile design variability.}

\begin{figure*}[htbp]
\centering
\includegraphics[width=\textwidth]{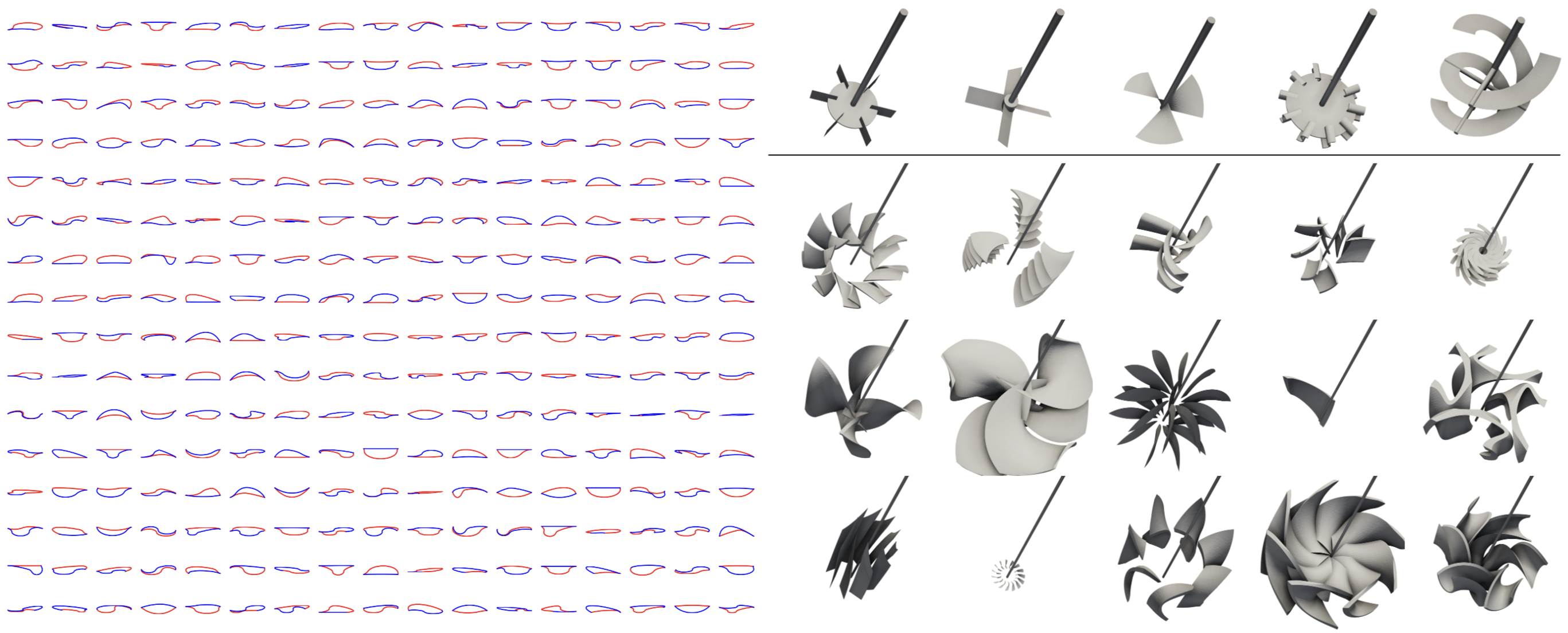}
\caption{ 
\textbf{Left:} examples of blade sections generated using only four free parameters defining the suction and pressure sides. The kernel spans a wide variety of shapes, including symmetric, cambered, open, C-curved and S-curved profiles, demonstrating its expressive capacity despite the low dimensionality. \textbf{Right:} examples of impellers generated from our geometric descriptors; \textit{top row}: standard industrial designs produced by manually selecting CAD kernel parameters; \textit{below}: examples of designs generated during the 1500 trials conducted in this study.}
\label{fig:profiles}
\end{figure*}

A key requirement of the HARPPP geometric kernel is to span a broad and physically relevant space of blade profiles while retaining a small number of interpretable parameters. Using only four free parameters to form the suction and pressure sides, the kernel is able to generate a remarkably wide variety of profiles (Fig.~\ref{fig:profiles}). 

The resulting design space includes thin and thick sections, open and closed contours, cambered and symmetric shapes, and both C- and S-curved profiles. This diversity is achieved without requiring explicit angular control parameters, since the symmetry properties of the generator allow inverted forms to arise naturally. Importantly, the kernel is not restricted to aerodynamic airfoil families, which enforce leading-edge smoothness and attached-flow conditions that are not relevant in mixing applications. Instead, the profiles can exhibit sharp curvature and non-monotonic camber, which HARPPP may exploit to tailor local turbulent dissipation.

The examples shown illustrate the breadth of possible forms accessible with only four degrees of freedom. Additional higher-order controls (e.g.\ stretch, twist, helix, repeats) then act upon these base profiles to generate the full impeller families explored in the optimisation study.

\paragraph{Implementation in this work.}
In the present study we restricted attention to thin-blade configurations, as these are most representative of the industrial designs provided in the benchmark brief. Accordingly, the full profile parametrisation was reduced to a single curve defined by only two parameters. This simplification specifies the leading-edge position and the extent of the profile while still allowing for substantial variability in curvature. The resulting two-parameter profile space was therefore sufficient for exploring a diverse yet computationally tractable range of blade designs in the optimisation study. Importantly, this reduction does not limit the generality of the kernel: the full four-parameter formulation remains available for future studies where thicker or multi-surface blades are of interest.

\subsubsection*{Blade Kernel}

The HARPPP framework incorporates a custom, directly-programmed CAD kernel, to generate fully parametric blade, impeller, and baffle geometries. This kernel translates the abstract descriptors defined in the optimisation problem into watertight solid models suitable for automated meshing. By integrating CAD generation directly into the simulation loop, we avoided the need for external, manual geometry preparation.

The geometric kernel exposes a set of intuitive parameters (Table~\ref{tab:impeller-params}) that control placement, aspect ratios, root and tip positions, higher-order deformations (twist, helix, curl), and repetition patterns. Baffles are described by a subset of these descriptors, reflecting their static rather than rotating role. Together, these parameters are sufficiently expressive to reproduce virtually all standard industrial stirred-tank designs, including Rushton, pitched-blade, propeller, high-shear, ribbon, and others, as special cases corresponding to particular parameter values. At the same time, the kernel generalises these conventional families to a continuous design space amenable to evolutionary search.

\begin{table}[htbp]
\centering
\caption{Programmed CAD kernel parameters from an example HARPPP blade geometric kernel. Baffles use a subset of these descriptors. The values used to reproduce the baseline (Rushton-6) design are included.}
\label{tab:impeller-params}
\footnotesize
\begin{tabular}{lccc}
\hline
\textbf{Parameter} & \textbf{Min} & \textbf{Max} & \textbf{Baseline} \\
\hline
Bottom clearance & 0 & 0.8 & 0.3 \\
Lateral $x$ placement & $-0.8$ & 0.8 & 0 \\
Lateral $y$ placement & $-0.8$ & 0.8 & 0 \\
Impeller height ratio & 0 & 0.8 & 0.092857 \\
Impeller diameter ratio & 0 & 0.8 & 0.335 \\
Blade root position & 0 & 1 & 0.597 \\
Blade extent in $x$ & 0 & 1 & 0.806 \\
Blade extent in $y$ & 0 & 1 & 0.003 \\
Joint place (relative) & 0 & 1 & 0.5 \\
Joint angle ($^\circ$) & $-180^\circ$ & $180^\circ$ & $0^\circ$ \\
Lean place (relative) & 0 & 1 & 0.5 \\
Lean angle ($^\circ$) & $-180^\circ$ & $180^\circ$ & $0^\circ$ \\
Turn place (relative) & 0 & 1 & 0.5 \\
Turn angle ($^\circ$) & $-180^\circ$ & $180^\circ$ & $0^\circ$ \\
Twist ($^\circ$) & $0^\circ$ & $360^\circ$ & $0^\circ$ \\
Helix ($^\circ$) & $0^\circ$ & $360^\circ$ & $0^\circ$ \\
Curl (mm) & 0 & 500 & 0 \\
Number of blades & 0.5 & 16.5 & 6 \\
Number of axial repeats & 0.5 & 8.5 & 1 \\
Repeats axial spacing & 0 & 4 & - \\
Repeats angular bias ($^\circ$) & $0^\circ$ & $180^\circ$ & - \\
Profile parameter $\alpha_1$ & 0 & 8 & 0 \\
Profile parameter $\beta_1$ & 0 & 5 & 0.01 \\
\hline
\end{tabular}
\end{table}

\subsection*{CFD simulation}

All candidate geometries generated by the CAD kernel were evaluated by three-dimensional transient CFD simulations embedded directly within the HARPPP optimisation loop. The workflow comprised automated meshing, solver setup, power-controlled transient integration, and post-processing of dissipation-based objectives.

\subsubsection*{Meshing} 
Each geometry (vessel, impeller, and baffles) was triangulated into STL surfaces and combined into a conformal computational domain. Initial grids were generated by a custom templated \texttt{blockMeshDict}, producing structured hexahedral blocks with radially graded resolution about the shaft. Surface snapping and local refinement were then applied using \texttt{snappyHexMesh}, with explicit feature refinement on the impeller blades, vessel walls, and baffles, and optional cylindrical refinement zones in the swept impeller region.

This procedure yielded watertight polyhedral meshes with conformal AMI interfaces between the rotating and stationary regions, suitable for transient CFD analysis. All meshing steps were fully scripted, ensuring reproducibility for every candidate geometry evaluated within the optimisation loop.

\subsubsection*{Solver setup}

All CFD simulations were performed using \texttt{pimpleFoam} (OpenFOAM v2312), with the PIMPLE algorithm (blended SIMPLE-PISO) to handle the transient, incompressible Navier-Stokes equations. Second-order backward differencing was used for time integration, and spatial discretisation employed linear-upwind schemes for convective terms with limited-corrected gradients and Laplacians. Pressure was solved using a GAMG multigrid algorithm, while velocity and turbulence quantities used smooth solvers with strict tolerances ($10^{-6}$-$10^{-8}$). Two outer corrector loops were applied within the PIMPLE algorithm to ensure strong coupling between the AMI-driven impeller motion and the resolved flow field.

Turbulence was modelled using the RANS $k$-$\omega$ SST closure, which provides reliable predictions of impeller-driven tank flows with moderate computational expense \cite{singh2011assessment}. Initial conditions for $k$, $\omega$, and $\nu_t$ were computed automatically from a reference length scale ($L_\mathrm{ref}$, equal to the impeller diameter) and a turbulence intensity of 1\%. No-slip conditions were imposed on vessel walls, baffles, and impeller blades, while the impeller shaft had a fixed rotation speed applied. The impeller region was handled with an Arbitrary Mesh Interface (AMI) to couple the rotating and stationary domains.

\subsubsection*{Temporal integration}

The baseline timestep was set to $\Delta t = 0.0015$\,s, corresponding to a resolution of $<1^{\circ}$ per impeller revolution at $\sim 100$\,rpm. Independence tests with $\Delta t=0.001$ and $0.002$\,s confirmed $<3\%$ variation in time-averaged quantities. Each candidate geometry was simulated for at least 30\,s, sufficient for initial transients to decay and statistics to stabilise (see Figure \ref{fig:validation}).

\subsubsection*{Power-control feedback}

To guarantee comparability across different geometries, all simulations were run at a constant target \emph{shaft} power of 3024\,W. This setpoint equals the shaft power of the Johnson Matthey (JM) \emph{reference} configuration (six-blade Rushton + four baffles) at its nominal operating point, as supplied from JM commissioning measurements for the benchmark vessel; expressed per volume this is \(P/V \approx 0.69~\mathrm{kW\,m^{-3}}\) for our \(4.40~\mathrm{m^{3}}\) charge. A proportional model-inversion feedback controller, implemented directly in the simulation loop, adjusted the impeller rotation rate at regular intervals (2.0\,s) by comparing the instantaneous power draw against the target. Power was calculated as torque integrated over the impeller surfaces multiplied by angular velocity, \(P(t)=\Omega(t)\int_{\mathcal{S}_\mathrm{imp}} \boldsymbol{r}\times\boldsymbol{f}\,\mathrm{d}S\). The controller updated \(\Omega(t)\) smoothly through tabulated ramps, ensuring stability while converging toward the prescribed power. Geometries that failed to complete the full 30\,s run were automatically marked invalid and excluded from optimisation.

\subsubsection*{Parallelisation and execution}

All cases were decomposed into MPI subdomains (in this study set to 32) using \texttt{scotch} partitioning and executed in parallel. Simulations were checkpointed automatically at each power-control interval, allowing failed runs to be detected and terminated without affecting subsequent optimisation.

\subsubsection*{Post-processing and objectives}

Post-processing was fully scripted within the HARPPP workflow. At each timestep, instantaneous forces, torque, and power were recorded from the impeller and shaft surfaces. Volumetric turbulence quantities ($\varepsilon$, $k$, $\omega$) were sampled and averaged over the final 15\,s of each run. The primary optimisation objectives--mean turbulent dissipation and its coefficient of variation--were derived from these statistics.

\subsection*{Optimisation}

\subsubsection*{ACCES: evolution strategy with in-loop parallel sampling}
The CAD-CFD pipeline was optimised using ACCES (\emph{Autonomous Characterisation and Calibration via Evolutionary Simulation Software}), a Python-based framework we developed for black-box simulation optimisation that directly launches parallel HPC workflows rather than requiring a wrapped objective function \cite{nicusan2025acces}.

Beyond adopting CMA-ES sampling and rank updates, ACCES contributes three extensions tailored to simulation-driven design. 
First, it implements \emph{in-loop, generation-level parallel orchestration} of CAD-CFD evaluations: each generation is materialised via metaprogramming (abstract syntax tree modification of Python scripts which drive a single simulation trial), dispatched in parallel, and synchronised at a barrier before adaptation. This removes the need to wrap the simulator behind a single black-box callable and permits scheduling on local OS processes or distributed clusters; in this work we used SLURM scheduling \cite{yoo2003slurm}. 
Second, \emph{phenotype shaping} is enforced by affine normalisation of all descriptors to a common unit hypercube and by choosing $\sigma^{(0)}$ so that the initial population spans $\approx 40\%$ of each dimension, yielding scale-consistent mutations across heterogeneous units (length ratios, angles, millimetres) and avoiding premature collapse in poorly scaled directions; this makes the optimiser completely hyperparameter-free. 
Third, \emph{evaluation censoring and archiving} are handled natively: numerically unstable simulations (e.g.\ runs not reaching the prescribed physical time) are treated as missing and excluded from selection, while all successful evaluations--together with their full case descriptors--are archived. This enables post hoc reconstruction of dissipation-uniformity trade-offs even when a scalarised objective is used in-loop. 
Together, these additions preserve the invariance and exploration properties of CMA\textendash ES while making the algorithm operationally fit for large-batch, failure-prone CFD pipelines.

\subsubsection*{Design Exploration via Evolutionary Optimisation}

At generation $g$, CMA--ES samples $\lambda$ candidates
$x_i^{(g)} \sim \mathcal{N}\!\left(m^{(g)},\,\sigma^{(g)2} C^{(g)}\right)$
and updates the sampling distribution by rank-based selection and adaptation \cite{hansen2016cma}. 
Let $x_{i:\lambda}^{(g)}$ denote the $i$th best sample (by fitness), $w_i>0$ normalised with $\sum_i w_i=1$, and 
$y_{i:\lambda}^{(g)}=\big(x_{i:\lambda}^{(g)}-m^{(g)}\big)/\sigma^{(g)}$.
The mean (recombination) update is
\[
m^{(g+1)} \;=\; \sum_{i=1}^{\mu} w_i\, x_{i:\lambda}^{(g)}\,,
\]
which is invariant to strictly monotone transformations of the objective and to affine re-scalings of the search space, mitigating parameterisation-induced bias in exploration \cite{hansen2016cma,hansen2001derandomized}.

Two complementary covariance updates encode exploration along successful directions. 
The \emph{rank-$\mu$} update accumulates information from the top $\mu$ samples
\[
C^{(g+1)} \;\leftarrow\; (1-c_1-c_\mu)\,C^{(g)} \;+\; c_\mu \sum_{i=1}^{\mu} w_i\, y_{i:\lambda}^{(g)} \,y_{i:\lambda}^{(g)\top} \,,
\]
while the \emph{rank-one} update exploits the evolution path $p_c^{(g+1)}$ (a running average of successful moves),
\[
\begin{aligned}
p_c^{(g+1)}
  &= (1-c_c)\,p_c^{(g)}
\\&\quad
   + h_\sigma\,\sqrt{c_c(2-c_c)\,\mu_w}\,
     \frac{m^{(g+1)}-m^{(g)}}{\sigma^{(g)}}\,,
\\[4pt]
C^{(g+1)}
  &\leftarrow C^{(g+1)} + c_1\, p_c^{(g+1)} p_c^{(g+1)\top}.
\end{aligned}
\]
with $\mu_w=\big(\sum_i w_i^2\big)^{-1}$ and indicator $h_\sigma$ \cite{hansen2016cma}.
The rank-$\mu$ term broadens $C$ along directions that repeatedly yield improvements, while the rank-one term elongates the distribution along the serial correlation of steps (the ``evolution path''), enabling discovery of extended high-performing manifolds (design families) rather than collapsing to a single point.

Step size is adapted via cumulative step-length control,
\[
\begin{aligned}
p_\sigma^{(g+1)}
  &= (1-c_\sigma)\,p_\sigma^{(g)}
\\&\quad
   + \sqrt{c_\sigma(2-c_\sigma)\,\mu_w}\;
     C^{(g)^{-1/2}}
     \frac{m^{(g+1)}-m^{(g)}}{\sigma^{(g)}}\,,
\\[4pt]
\sigma^{(g+1)}
  &= \sigma^{(g)} \exp\!\left(
       \frac{c_\sigma}{d_\sigma}
       \left(
         \frac{\|p_\sigma^{(g+1)}\|}{\mathbb{E}\,\|\mathcal{N}(0,I)\|}
         - 1
       \right)
     \right).
\end{aligned}
\]
which expands $\sigma$ when progress directions decorrelate (flat/noisy regions) and contracts it under consistent progress, yielding scale-free alternation between global probing and local refinement without ad hoc schedules \cite{hansen2016cma,hansen2001derandomized}. 

In the information-geometric optimisation (IGO) view, these rank-based, invariant updates implement a natural-gradient ascent on a quantile-weighted objective in distribution space, moving the \emph{entire} sampling distribution toward better regions with minimal Kullback-Leibler change per iteration -- thus preserving diversity and supporting exploration \cite{ollivier2017information}.

\subsubsection*{Optimisation budget and reproducibility.}
The impeller search used \(\lambda=50\) candidates per generation for 30 generations
(\(N=1{,}500\) CFD evaluations). All parameters were affine-scaled to similar ranges with unit variance, and the
initial step size was set so that the first population spanned \(\approx 40\%\) of each
dimension (23-dimensional space). Failed runs (numerical instability or early stop)
were censored (NaN) and not used for selection. The full per-generation
archives are provided in Supplementary Information.

\backmatter

\section*{Acknowledgements}

The computations described in this paper were performed using the University of Birmingham's BlueBEAR HPC service, which provides a High Performance Computing service to the University's research community. See http://www.birmingham.ac.uk/bear for more details.

We also gratefully acknowledge the support of the EPSRC Centre for Doctoral Training in Formulation Engineering for Net Zero at the University of Birmingham.

\section*{Author contributions}

\textbf{Andrei-Leonard Nicușan (ALN)}: project conceptualisation, project administration, investigation, methodology, software, validation, visualisation, writing - original draft preparation. \textbf{Darren Gobby (DG)}: project conceptualisation, project administration, writing - review. \textbf{Kit Windows-Yule (KWY)}: project conceptualisation, project administration, resources, writing - review \& editing.

\section*{Competing interests}

The authors declare no competing interests.

\section*{Supplementary information}

The optimisation artefacts (identified near-front designs and full per-generation logs) are available at \url{https://github.com/anicusan/HARPPP-CSTR-Designs}, containing:
\begin{itemize}
\item \textbf{Supplementary Data~1 (impellers)}: STL triangulations of the selected optimised designs.
\item \textbf{Supplementary Data~2 (baffles)}: VTP triangulations of the selected optimised designs.
\item \textbf{Supplementary Data~3 (optimisation)}: full optimisation log and running script; data format explained in the readme.rst file.
\end{itemize}

%
%
%

\bibliography{sn-bibliography}

\end{document}